\documentclass[fleqn,twoside]{article}
\usepackage{espcrc2}

\usepackage{graphicx}
\usepackage[figuresright]{rotating}

\newcommand{\AmS}{{\protect\the\textfont2
  A\kern-.1667em\lower.5ex\hbox{M}\kern-.125emS}}

\title{Supernova neutrino challenges}

\author{C. Y. Cardall\address{Physics Division, Oak Ridge National Laboratory, \\
      Oak Ridge, TN 37831-6354, United States of America}%
      \address{Department of Physics and Astronomy, University of Tennessee, \\
	Knoxville, TN 37996-1200, United States of America}%
        \thanks{This work was supported 
by  Scientific Discovery Through
Advanced Computing (SciDAC), a program of the Office of Science of the U.S. Department of Energy (DoE); and by Oak Ridge National Laboratory, managed by UT-Battelle, LLC, for the DoE under contract DE-AC05-00OR22725.}}
       
\begin{document}


\begin{abstract}
A principal `supernova neutrino challenge' is the computational difficulty of six-dimensional neutrino radiation hydrodynamics. The variety of resulting approximations has provoked a long history of uncertainty in the core-collapse supernova explosion mechanism, but recent work highlighting low-mode convection and a newly-recognized  instability in spherical accretion shocks may signal (yet another) resolution. As part of its goal of elucidating the explosion mechanism, the Terascale Supernova Initiative is committed to meeting the full complexity of the computational challenge. The understanding of supernova neutrino emission gained in detailed simulations provides a potential basis for learning about two major remaining unknowns in neutrino flavor mixing: the value of the mixing angle $\theta_{13}$, and distinguishing between ``normal'' and ``inverted'' mass hierarchies.
\vspace{1pc}
\end{abstract}

\maketitle

\section{CORE-COLLAPSE SUPERNOVAE}

Shortly after the discovery of the neutron in the early 1930s, Baade and Zwicky declared, 
``With all reserve we advance the view that supernovae represent the
transitions from ordinary stars to {\em neutron stars,} which in their final
stages consist of extremely closely packed neutrons'' \cite{baade34}. 
This turned out to be true, at least for some `core-collapse' supernovae (those of Type Ib/Ic/II); a black hole is another possible outcome. 
The dominant fleshing-out of the core collapse process in the last two decades
has been
the delayed neutrino-driven explosion mechanism
\cite{wilson85,bethe85}.

A core-collapse supernova results from the evolution of a massive star.
For most of their existence, stars burn hydrogen into helium. 
In stars at least eight times as massive as the Sun ($8\ M_\odot$), 
temperatures and densities become sufficiently high to burn 
through carbon to oxygen, neon, and magnesium; in stars of at least $\sim 10\ M_\odot$, burning continues through silicon to iron group elements. 
The iron group nuclei are the most tightly bound, and here burning in 
the core ceases. 

The iron core---supported by electron 
degeneracy pressure instead of gas thermal pressure, because of cooling
by neutrino emission from carbon burning onwards---eventually 
becomes unstable. Its inner portion undergoes homologous collapse
(velocity proportional to radius), and the outer portion collapses 
supersonically. 
Electron capture on nuclei is one instability leading to collapse, and 
this process continues throughout collapse, producing neutrinos.
These neutrinos escape freely
until densities 
in the collapsing core
become so high that even neutrinos are 
trapped. 

Collapse is halted soon after the matter exceeds nuclear density; 
at this point (``bounce''), a shock wave forms at the boundary between the homologous
and supersonically collapsing regions. The shock begins to move out,
but after the shock passes
some distance beyond the surface of the newly-born neutron star, 
it stalls as energy 
is lost to neutrino emission and endothermic dissociation of heavy 
nuclei falling through the shock.

It is natural to consider neutrino heating as a mechanism for
shock revival, because neutrinos dominate the energetics of
the post-bounce evolution.
Initially, the nascent neutron star is a hot thermal bath of dense nuclear matter, 
electron/positron pairs, photons, and neutrinos, containing most of 
the gravitational potential energy released during core collapse. 
Neutrinos, having the weakest interactions, are the most efficient 
means of cooling; they diffuse outward on a time scale of seconds, 
and eventually escape with about 99\% of the released gravitational energy.

Because neutrinos dominate the energetics of the system, 
a detailed understanding of their evolution will be integral to 
definitive accounts of the supernova process.
If we want to understand the explosion---which accounts for only about 1\% of 
the energy budget of the system---we should carefully account for the
neutrinos' much larger contribution to the energy budget.

What sort of computation is needed to follow the neutrinos' evolution?
 Deep inside the newly-born neutron star, 
 the neutrinos and the fluid are tightly coupled (nearly in equilibrium);  but as 
 neutrinos are transported from inside the neutron star, they go from a nearly isotropic diffusive regime to strongly forward-peaked free-streaming. Heating 
 behind the shock occurs precisely in 
this transition region, and modeling this process accurately requires tracking both the energy and angle dependence of the neutrino distribution functions at every point
in space. 

A full treatment of this six-dimensional neutrino radiation hydrodynamics
problem is a major challenge, 
too costly for 
contemporary
computational resources. While much has been 
learned over the years through simulation of model systems of reduced dimensionality, there is as yet no robust confirmation of the delayed neutrino-driven scenario described above (see Sec. \ref{sec:history}).

Of interest are recent detections of a handful of unusually energetic Type Ib/c supernovae (often called `hypernovae') in connection with gamma-ray bursts. 
Some show evidence for jets, and some do not. 
Determining the relative rates of jet-like hypernovae, non--jet-like hypernovae, and `normal' supernovae,
and the possible associated variety of mechanisms, are 
important challenges.

In summary, the details of how the stalled shock is revived
sufficiently to continue plowing through the outer layers of the
progenitor star are unclear. In normal supernovae, it may well be that some combination of neutrino heating 
behind the shock, convection, and instability of the spherical
accretion shock leads to the explosion (see Sec. \ref{sec:history}). It is tempting to think that rotation 
and magnetic fields 
may play a more significant role in the rare jet-like hypernovae, perhaps giving birth to `magnetars,' the class of neutron stars with unusually large magnetic fields. 

\section{HISTORY OF NEUTRINO RADIATION HYDRODYNAMICS}
\label{sec:history}

From the description of the core-collapse supernova process in the
previous section, several key aspects of physics that a simulation must 
address can be identified: gravity, fluid dynamics (including equation of state and composition), neutrino transport and interactions, and perhaps magnetic fields. Ideally each of these aspects would be implemented in general relativity. 

Here the last two decades' progress on just one of these simulation aspects is sketched: the high dimensionality (three space and three momentum space dimensions---not to mention time dependence) of neutrino radiation hydrodynamics (see Table 1). The development of this aspect of the simulations is intertwined with important advances in the field, but of course does not represent every insight relevant to the explosion mechanism obtained via simulation or otherwise. 

\begin{table*}
\caption{Selected neutrino radiation hydrodynamics milestones in stellar collapse  simulations studying the long-term fate of the shock.
The ``Yes'' entries in the  ``Explosion'' column are all marked with an asterisk as a reminder that questions about the simulations---described in the main text---have prevented a consensus about the explosion mechanism. 
``Total dimensions'' is the average of ``Fluid space dimensions'' and  ``$\nu$ space dimensions,'' added to ``$\nu$ momentum dimensions.''
The abbreviation ``N'' stands for `Newtonian,' while ``PN''---for `Post-Newtonian'---stands for some attempt at inclusion of general relativistic effects, and ``GR'' denotes full relativity. A space dimensionality in quotes---like ``1.5''---denotes an attempt at modeling higher dimensional effects within the context of a lower dimensional simulation. For the fluid, this is a mixing-length prescription in the neutron star (``NS'') or the heating region (``HR'') behind the stalled shock. For neutrino transport, it indicates one of two approaches: multidimensional diffusion in regions with strong radiation/fluid coupling, matched with a spherically symmetric `light bulb' approximation in weakly coupled regions (``thick/thin''); or the (mostly) independent application of a spherically symmetric formalism/algorithm to separate spatial angle bins (``ray-by-ray'').
}
\begin{tabular}{@{}lcccccc}
\hline
Group & Year & Explosion & Total  & Fluid space  & $\nu$ space & $\nu$ momentum
	     \\
	     &       & & dimensions    & dimensions & dimensions & space dimensions \\
\hline
Lawrence & 1982 & Yes$^*$ & 2 & 1  & 1 & 1    \\
 Livermore \cite{bowers82,wilson85,bethe85} & & & & (PN)  & &(${\cal O}(v/c)$)  \\
\hline
Lawrence & 1985 & Yes$^*$ & ``2.25'' & ``1.5'' NS  & 1 & 1  \\
 Livermore \cite{mayle85,wilson88} & & & & (PN)  & &(${\cal O}(v/c)$)  \\
\hline
Florida & 1987 & No & 2 & 1  & 1 & 1  \\
Atlantic \cite{bruenn85,bruenn87,bruenn91,bruenn93}  & & & & (GR)  & &(${\cal O}(v/c)$) \\
\hline
Lawrence & 1989 & Yes$^*$ & ``2.25'' & ``1.5'' NS+HR  & 1 & 1   \\
 Livermore \cite{mayle90,mayle91,wilson93} & & & & (GR)  & & (GR) \\
\hline
Lawrence & 1992 & Yes$^*$ & 2 & 2 HR  & 2 & 0  \\
 Livermore \cite{miller93} & & & & (N)  & &(N) \\
\hline
Los  Alamos & 1993  & Yes$^*$ & ``1.75'' & 2  & ``1.5'' & 0 \\
\cite{herant94}   & & & & (N)  & thick/thin &(PN) \\
Arizona & 1994 & Yes$^*$ & ``1.75'' & 2  & ``1.5'' & 0  \\
\cite{burrows95}  &  & & & (N)  & ray-by-ray & (N) \\
\hline
Florida & 1994  & No & ``2.25'' & ``1.5'' NS & 1 & 1 \\
Atlantic \cite{bruenn94,bruenn95} & & & & (GR)  & &(${\cal O}(v/c)$) \\
\hline
Oak Ridge & 1996 & No & ``2.5'' & 2 & 1 & 1  \\
\cite{mezzacappa98a,mezzacappa98b} & & & & (N)  & &(${\cal O}(v/c)$) \\
\hline
Max Planck & 2000 & No, Yes$^*$ & 3 & 1 & 1 & 2  \\
\cite{rampp00,rampp02,kitaura03,janka04b} & &  (ONeMg)& & (N)  & &(${\cal O}(v/c)$) \\
Oak Ridge & 2000 & No & 3 & 1 & 1 & 2  \\
\cite{mezzacappa93b,mezzacappa93c,mezzacappa99,liebendoerfer00,mezzacappa01} & & & & (N)  & &(${\cal O}(v/c)$) \\
Arizona & 2002 & No & 3 & 1 & 1 & 2  \\
\cite{burrows00,thompson03} & & & & (N)  & &(${\cal O}(v/c)$) \\
\hline
Oak Ridge & 2000  & No & 3 & 1 & 1 & 2 \\
\cite{mezzacappa93b,mezzacappa93c,mezzacappa99,liebendoerfer01a,liebendoerfer02,liebendoerfer04} & & & & (GR)  & &(GR) \\
\hline
Los  Alamos & 2002 & Yes$^*$ & ``2.5'' & 3  & ``2'' & 0  \\
\cite{herant94,fryer02}  &  & & & (N)  & thick/thin &(PN) \\
\hline
Max Planck & 2003 & No, Yes$^*$ & ``3.75'' & 2 & ``1.5'' & 2 (${\cal O}(v/c)$,  \\
\cite{rampp02,buras03,janka02,janka04a,janka04b} & & (180$^{\mathrm{o}}$) & & (PN)  & ray-by-ray & PN) \\
\hline
\end{tabular}\\[2pt]
\label{history}
\end{table*}

I pick up the story in 1982, when simulations showing the stalled shock reenergized by neutrino heating on a time scale of hundreds of milliseconds were first performed \cite{wilson85}. This was initially achieved in a simulation with a total of 2 dimensions (spherical symmetry, and energy-dependent neutrino transport). But 
with the introduction of full general relativity and a correction in an outer boundary condition \cite{mayle90}, it became clear that these models would not explode without a mock-up of a doubly-diffusive fluid instability in the newly-born neutron star that serves to boost neutrino luminosities \cite{mayle90,mayle91,wilson93,miller93}---a simulation of effective total dimensionality ``2.5'' (see Table 1). That the necessary conditions exist for this particular instability to operate has been disputed \cite{bruenn95,bruenn96,bruenn04}; and though related phenomena may operate \cite{bruenn04}, more recent simulations with energy-dependent neutrino transport and true two-dimensional fluid dynamics indicate that fluid motions are either suppressed by neutrino transport \cite{mezzacappa98a} or have little effect on neutrino luminosities and supernova dynamics \cite{buras03,janka04a}.

Recognizing that the profiles obtained in spherical symmetry implied convective instabilities, and that observations of supernova 1987A also pointed to asphericities, several groups explored fluid motions in two spatial dimensions in the supernova environment in the 1990s. 

One class of simplifications allowed for neutrino transport in ``1.5'' or 2 spatial dimensions, but with
neutrino energy and angle dependence integrated out, 
reducing a five dimensional problem to  ``1.75'' or 2 effective total dimensions (see Table 1) \cite{miller93,herant94,burrows95}.
These simulations exhibited explosions, and elucidated an undeniably important physical effect: a negative entropy gradient behind the stalled shock results in convection that increases the efficiency of heating by neutrinos.
However, in the scheme of Table 1, the inability to track the neutrino energy dependence in these simulations could be viewed as a minor step backwards in effective total dimensionality. The energy dependence of neutrino interactions has the important effect of enhancing core deleptonization, which makes explosions more difficult \cite{bruenn85,bruenn89a,bruenn89b}; this raised the question of whether the exploding models of the early- and mid-1990s were too optimistic. 

This concern about the lack of neutrino energy dependence received some support from a simulation in the late 1990s involving 
a different simplification of neutrino transport:
the imposition of energy-dependent
neutrino distributions from spherically symmetric simulations
onto fluid dynamics in two spatial dimensions
\cite{mezzacappa98b}. 
Unlike the 
simulations
discussed above, these did not 
explode, casting doubt
upon claims that convection-aided
neutrino heating constituted a robust explosion mechanism.

The nagging qualitative difference between spatially multidimensional
simulations with different neutrino transport approximations 
motivated interest in the possible importance of even more complete neutrino
transport: Might the retention of both the energy
{\em and} angle dependence of the neutrino distributions improve the chances of explosion, as preliminary ``snapshot'' studies suggested \cite{messer98,burrows00}?
Of necessity, the first such simulations were performed in
spherical symmetry, which nevertheless represented an advance to a total dimensionality of 3 (see Table 1).
Results from
three different groups are in accord: Spherically symmetric models of iron core collapse 
do not explode, even with solid neutrino transport 
\cite{rampp00,mezzacappa01,thompson03} and general relativity \cite{liebendoerfer01a,liebendoerfer04}. Recently, however, it has been shown that the more modest oxygen/neon/magnesium cores of the lightest stars to undergo core collapse (8-10 M$_\odot$) may explode in spherical symmetry \cite{kitaura03,janka04b}.

The current state of the art in neutrino transport in supernova simulations determining the long-term fate of the shock has been achieved by one of the above groups (centered at the Max Planck Institute for Astrophysics in Garching), who deployed their spherically symmetric energy- and angle-dependent neutrino transport capability \cite{rampp02} along separate radial rays, with partial coupling between rays \cite{janka02}. Initial results---from axisymmetric simulations with a restricted angular domain---were negative with regards to explosions (in spite of the salutary effects of convection, and also rotation) \cite{buras03}, apparently supporting the results of Ref. \cite{mezzacappa98b}. 
An explosion was seen in one simulation \cite{janka03}
in which
certain terms in the neutrino transport equation corresponding to Doppler shifts
and angular aberration due to fluid motion were dropped; this simulation also yielded a neutron star mass and nucleosynthetic consequences in better agreement with observations than the ``successful'' explosion simulations of the 1990s \cite{herant94,burrows95}, arguably because of more accurate neutrino transport in the case of both observables. 
The continuing lesson
is that getting the details of the neutrino transport right makes a difference.

In addition to accurate neutrino transport, low-mode ($\ell = 1,2$) instabilities that can develop only in simulations allowing the full range of polar angles may make a subtle but decisive difference, as in an explosion recently reported by the Garching group \cite{janka04a,janka04b}. This achievement was presaged by earlier studies demonstrating the tendency for convective cells to merge to the lowest order allowed by the spatial domain \cite{herant92} and recognizing a new spherical accretion shock instability \cite{blondin03} (discovered independently in a different context in Ref. \cite{foglizzo02}). These global asymmetries may even be sufficient to account for observed asphericities that have often been attributed to rotation and/or magnetic fields. 

Surely every `Yes' entry in the explosion column of Table 1 has been hailed in its time as `the answer' (at least by some!), and as a community we cannot help hoping once again that these recent developments mark the turning of a corner; but important work remains to verify if this is the case. Several groups are committed to further efforts. For example, the Terascale Supernova Initiative (TSI, which includes authors of Ref. \cite{blondin03}) comprises efforts aimed at ``ray-by-ray'' simulations \cite{hix01} like those of the Garching group, as well as full spatially multdimensional neutrino transport, both with energy dependence only \cite{myra04} and with energy {\em and} angle dependence \cite{cardall04}. Delineation of the possible roles of rotation and magnetic fields are also being pursued by TSI. At least one other group is pursuing full spatially multidimensional neutrino transport \cite{livne04,walder04}.








\section{SUPERNOVA NEUTRINOS AND FLAVOR MIXING}

As one who is engaged in the long-term development of a new code to be used in supernova simulations---with results from this code not yet in hand---I have spent this opportunity discussing our field's response to the challenge of neutrino radiation hydrodynamics in core-collapse supernova simulations over the last two decades, together with many references. Hopefully this provides some understanding of the history of claims in our field, and gives an appreciation of the years of dedicated work by many individuals to advance the state of the art. But in closing, I should at least say something more directly tied to the topic of the conference! 

So how do these large-scale simulations relate to neutrino flavor mixing? At present, the simulations assume massless neutrinos, relying on the standard expectation that the effective mass from neutrino forward scattering off electrons suppresses flavor mixing in the high-density region where the explosion is launched.\footnote{In public talks (presenting analyses not yet published), G.~M. Fuller has argued that the nonlinear effects of neutrino-neutrino forward scattering might produce flavor-changing effects even at the high densities near the nascent neutron star. If this proves correct, it would take the ``supernova neutrino challenge'' already faced by collapse modelers to a whole new level.} Before reaching terrestrial detectors, however, neutrinos will pass through flavor-changing resonances at lower densities in the stellar envelope (and possibly also experience flavor-changing effects while traversing Earth before entering a detector). Hence collapse simulations provide neutrino spectra as input to studies that take these effects into account in determining possible signals in terrestrial neutrino detectors (see, for example, overviews in Refs. \cite{mirizzi05,raffelt05}). To the extent that generic energy- and time-dependent features of the signals can be motivated by the large-scale simulations, such studies provide a possible basis for learning---from neutrino detection of a future Galactic supernova---about two major remaining unknowns surrounding neutrino flavor mixing: the value of the mixing angle $\theta_{13}$, and distinguishing between ``normal'' and ``inverted'' mass hierarchies. Alternatively, to the extent that these neutrino mixing features are determined by terrestrial long-baseline neutrino experiments before the next Galactic supernova, neutrino signals will allow the details of supernova dynamics to be observed in real time. 

\def\aap{Astron. Astrophys. }
\def\aasma{Am. Astron. Soc. Meet. Abs. }
\def\apj{Astrophys. J. }
\def\apjl{Astrophys. J. Lett. }
\def\apjs{Astrophys. J. Supp. Ser. }
\def\araa{Annu. Rev. Astron. Astrophys. }
\def\baas{Bull. Am. Astron. Soc. }
\def\jcam{J. Comp. Appl. Math. }
\def\nat{Nature }
\def\npa{Nucl. Phys. A }
\def\pr{Phys. Rep. }
\def\prd{Phys. Rev. D }
\def\prl{Phys. Rev. Lett. }
\def\prv{Phys. Rev. }


\end{document}